\documentclass[twocolumn]{aastex631}

\usepackage{amsmath}

\begin{document}

\title{Quenching and recovery of persistent X-ray emission during a superburst in 4U 1820--30}

\correspondingauthor{Zhaosheng Li, Yuanyue Pan}
\email{lizhaosheng@xtu.edu.cn, panyy@xtu.edu.cn}

\author{Zhijiao Peng}
\affiliation{Key Laboratory of Stars and Interstellar Medium, Xiangtan University, Xiangtan 411105, Hunan, P.R. China}

\author[0000-0003-3095-6065]{Zhaosheng Li}
\affiliation{Key Laboratory of Stars and Interstellar Medium, Xiangtan University, Xiangtan 411105, Hunan, P.R. China}

\author{Yuanyue Pan}
\affiliation{Key Laboratory of Stars and Interstellar Medium, Xiangtan University, Xiangtan 411105, Hunan, P.R. China}

\author{Tao Fu}
\affiliation{Key Laboratory of Stars and Interstellar Medium, Xiangtan University, Xiangtan 411105, Hunan, P.R. China}

\author{Wenhui Yu}
\affiliation{Key Laboratory of Stars and Interstellar Medium, Xiangtan University, Xiangtan 411105, Hunan, P.R. China}

\author[0000-0001-8768-3294]{Yupeng Chen}
\affiliation{Key Laboratory of Particle Astrophysics, Institute of High Energy Physics, Chinese Academy of Sciences, 19B Yuquan Road,
Beijing 100049, China}
\author{Shu Zhang}
\affiliation{Key Laboratory of Particle Astrophysics, Institute of High Energy Physics, Chinese Academy of Sciences, 19B Yuquan Road,
Beijing 100049, China}
\author[0000-0003-3095-6065]{Maurizio Falanga}
\affiliation{International Space Science Institute (ISSI), Hallerstrasse 6, 3012 Bern, Switzerland}
\affiliation{Physikalisches Institut, University of Bern, Sidlerstrasse 5, 3012 Bern, Switzerland}
\author[0000-0001-5586-1017]{Shuang-Nan Zhang}
\affiliation{Key Laboratory of Particle Astrophysics, Institute of High Energy Physics, Chinese Academy of Sciences, 19B Yuquan Road,
Beijing 100049, China}

\begin{abstract}

We report the superburst from 4U 1820--30 in 2021 observed by the Monitor of All-sky X-ray Image and Neutron star Interior Composition Explorer (NICER). During the tail of the superburst, we found that the NICER light curve unexpectedly increased from 1080 to 2204 ${\rm counts~s^{-1}}$ over 6.89 hr. From the time-resolved superburst spectra, we estimated the burst decay time of $\approx2.5$ hr, the ignition column depth of $\approx0.3\times 10^{12}~{\rm g ~cm^{-2}}$, the energy release per unit mass of $\approx2.4\times 10^{17}~{\rm erg~g^{-1}}$, the fluence of $\approx4.1\times 10^{-4}~{\rm erg~cm^{-2}}$, and the total energy release of $\approx3.5\times10^{42}$ erg. Notably, we found a gradual increase in the Componization flux from $8.9\times 10^{-10}~{\rm erg~s^{-1}~cm^{-2}}$ to the preburst level during the superburst. This increase can be interpreted as a consequence of superburst radiation depleting the inner accretion disk, leading to a near-complete quenching of the persistent emission. As the burst radiation decayed, the inner accretion disk gradually returned to its preburst state, as evidenced by the best-fit spectral parameters. Additionally, we observed a prominent absorption line that exhibited a gravitational redshift, shifting from 4.15 to 3.62 keV during the recovery phase of persistent emission. This absorption feature likely originates from the inner accretion disk rather than from burst emission on the neutron star (NS) surface. The observed changes in the absorption line energy suggest that the inner disk approached the NS to a distance as close as $\approx17$ km.

\end{abstract}


\section{Introduction}\label{sec:intro}

Neutron stars (NSs) in low-mass X-ray binary systems accrete matter from a Roche-lobe-overflowing, low-mass companion star \citep[$M<M_{\odot}$;][]{1992apa..book.....F}.
The accreted material is compressed and heated on the NS surface, which occasionally triggers unstable thermonuclear burning, known as type I X-ray bursts \citep[see][for reviews]{1993SSRv...62..223L,2006csxs.book..113S,2021ASSL..461..209G}.
Normal type I X-ray bursts are powered by helium or a mixture of hydrogen and helium, which have a rise time of a few seconds and usually last $\sim10\text{--}100$ s \citep[see][]{1993SSRv...62..223L,2008ApJS..179..360G,2020ApJS..249...32G}. 
Superbursts are thermonuclear shell flash fueled by carbon at an ignition column depth of $\sim10^{12}~{\rm g~cm^{-2}}$ \citep{2001ApJ...559L.127C,2002ApJ...566.1045S}. Superbursts usually last from hours to days, release a total energy of $\sim10^{42}\text{--}10^{43}$ erg, and recur from days to years \citep{2001ApJ...559L.127C,2017symm.conf..121I}. Observationally, all superbursts, except the one from 4U 0614+091, occurred with accretion rates higher than 10\% of the Eddington luminosity \citep{2017symm.conf..121I}.

Both normal type I X-ray burst and superburst spectra can usually be described as a diluted blackbody with a temperature of 0.5--3 keV and a radius of a few to thousand kilometers \citep{2008ApJS..179..360G,2024A&A...683A..93Y}. 
In some X-ray bursts, the peak fluxes can reach or slightly exceed the Eddington limit, as indicated by photospheric radius expansion (PRE) observed in time-resolved spectroscopy. 
The intense radiation of these bursts and the associated PRE process can lead to significant interactions between the burst emission and the surrounding accretion environment. 
Observations and numerical simulations suggest that these interactions are governed by several physical mechanisms \citep{2018SSRv..214...15D}. The radiation pressure of the burst emission drives an outflow \citep{Russell24}. X-ray heating may also alter the disk structure, likely increasing its scale height and changing the density \citep{2020NatAs...4..541F}. The Poynting–Robertson drag could remove angular momentum from the disk, enhancing mass accretion onto the NS surface \citep{Walker92,2020NatAs...4..541F,Zhao22,2023MNRAS.526.1388S}. Additionally, soft X-ray photons from the burst may cool the hot corona, leading to a reduction in hard X-ray emission during bursts \citep{Maccarone03,Chen18,2020MNRAS.499.4479S,Fu24}. The reflection of burst photons from the surrounding accretion disk has been observed during normal X-ray bursts, intermediate bursts, and superbursts \citep{Ballantyne04,Keek14,Zhao22,Lu24,2024A&A...683A..93Y}.

4U 1820--30 is a persistent atoll X-ray source, first identified by the Uhuru satellite \citep{1974ApJS...27...37G}. It is located in the globular cluster NGC 6642 at a distance of $8.4 \pm 0.6 $~kpc \citep{2004MNRAS.354..815V}. With a binary orbital period of 11.4~minutes, 4U 1820--30 is classified as an ultracompact X-ray binary \citep{1987ApJ...312L..17S}. 
4U 1820--30 were observed 15 type I X-ray bursts by the Neutron star Interior
Composition Explorer (NICER) during the 2017--2023 observations, and these bursts were powered by unstable thermonuclear burning of hydrogen-deficient material \citep{2024ApJ...975...67J,2024A&A...683A..93Y}. The first superburst from 4U 1820--30 was discovered by RXTE \citep{2002ApJ...566.1045S}. The superburst spectra showed a broad emission line between 5.8 and 6.4 keV and an absorption edge at 8--9 keV \citep{2002ApJ...566.1045S}. Alternatively, \citet{Ballantyne04} adopted a reflection model to fit the burst spectra and found that the superburst distorted the inner accretion disk.

Joint observations of NICER and the Monitor of All-sky X-ray Image (MAXI) allow us to detect and study the properties of long X-ray bursts \citep{Li21, Lu24}. In this paper, we present new observations of a superburst from 4U 1820--30, captured simultaneously by NICER and MAXI in 2021 August. The joint observations represent the most detailed soft X-ray spectroscopy and full temporal coverage of the superburst to date, which enables us to resolve the time-resolved evolution of the persistent emission and burst spectra. These data will also offer a unique opportunity to probe the disk's response to the intense radiation of the superburst and to test existing theoretical models of burst-disk interactions. In Sect.~\ref{sec:2}, we introduce the NICER and MAXI observations. In Sect.~\ref{sec:3}, we analyze the time-resolved persistent emission and burst spectra for the superburst. We discuss and summarize the results in Sections~\ref{sec:4} and \ref{sec:5}, respectively.

\section{OBSERVATION And DATA REDUCTION} \label{sec:2}

\begin{figure}[h]
    \centering
    \includegraphics[width=1\linewidth]{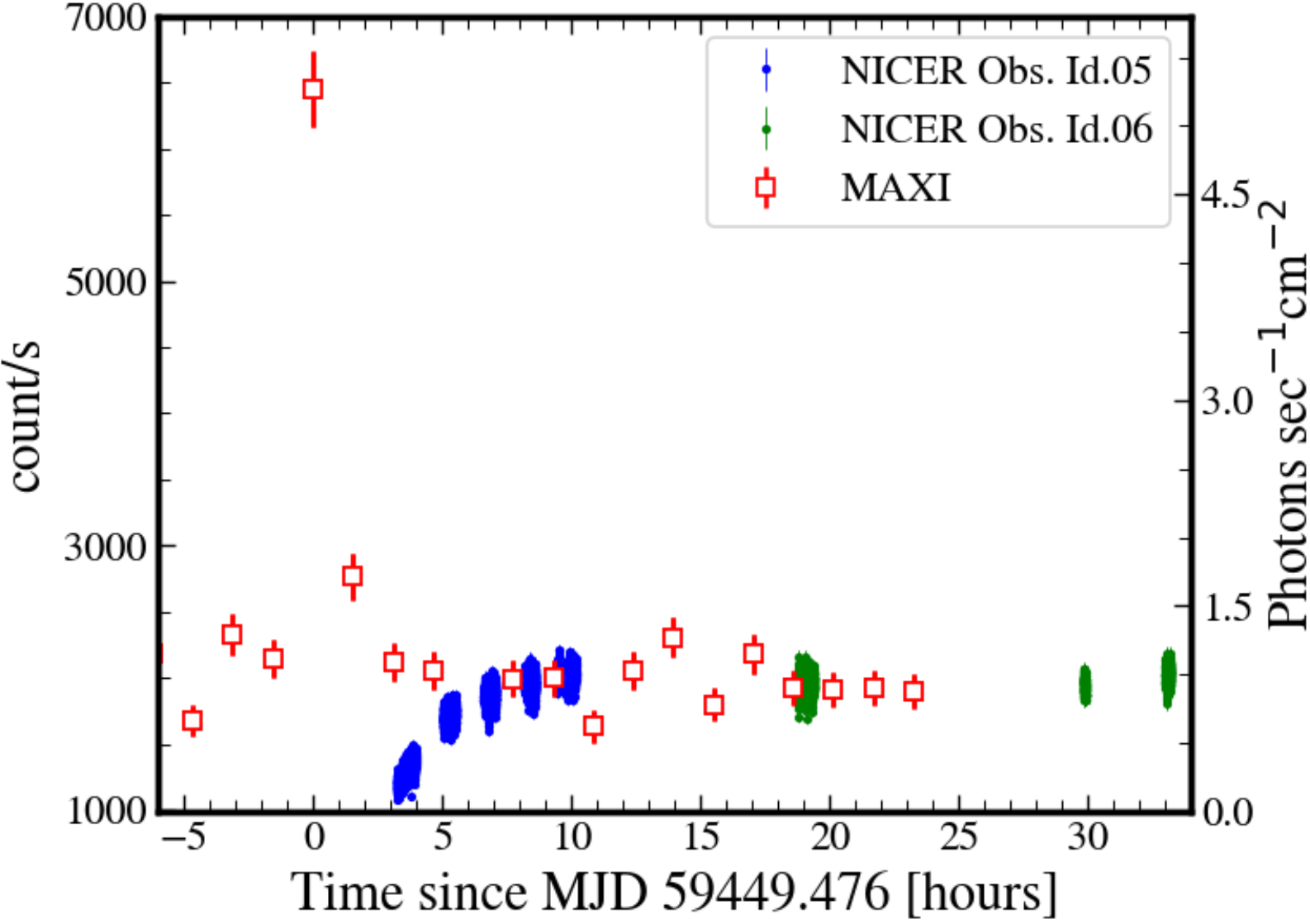}
    \caption{The light curve of the superburst from 4U~1820--30 observed by NICER (with a time bin size of 1 s in 0.5--10 keV; blue and green dots for ObsIDs 05 and 06, respectively) in units of $\rm counts~s^{-1}$ and MAXI ($\sim$250 s, 2--20 keV; open red squares) in units of $\rm{photons~sec^{-1}~cm^{-2}}$.}
    \label{fig:1}
\end{figure}

Two long X-ray bursts from 4U 1820--30 have been reported on the MAXI novae webpage, occurring on 2021 August 23 (MJD 59449.4768) and 2021 November 25 (MJD 59543.9764).\footnote{\href{http://maxi.riken.jp/alert/novae/}{http://maxi.riken.jp/alert/novae/}} In this work, we focus on the first long X-ray burst, as no NICER data are available for the second. We downloaded the 2--20 keV light curves with a bin size of 1.48~hr from MAXI/GSC\footnote{\href{http://maxi.riken.jp/mxondem/}{http://maxi.riken.jp/mxondem/}}; see Fig. \ref{fig:1}. 
The burst light curve is modeled using an exponential function, $C(t)=C(t_0)e^{-(t-t{_0})/\tau_{\rm LC}}+C_0$, where $C(t_0)$ is the normalization, $\tau_{\rm LC}$ is the exponential decay time, and $C_0$ accounts for the persistent count rate. The MAXI trigger time is adopted as the reference point for $t_0$. The model fit yields an exponential decay time of $\tau_{\rm LC} = 0.97 \pm 0.12$~hr, suggesting that this is a superburst.

The NICER archived data from 4U 1820--30 were searched for the period before and during the superburst.\footnote{No NICER data were available during the second long X-ray burst.} We found three observations, including ObsID 4663010102 (MJD 59385.10--59385.39), ObsID 4050300105 (MJD 59449.61--59449.90), and ObsID 4050300106 (MJD 59450.26--59450.86), starting at -64.5~days, 3.28~hr and 18.78~hr since the MAXI trigger, respectively, with a total unfiltered exposure time of 20.986~ks. Hereafter, we refer to these observations as ObsIDs 02, 05, and 06.

\begin{figure}[]
    \centering
    \includegraphics[width=1\linewidth]{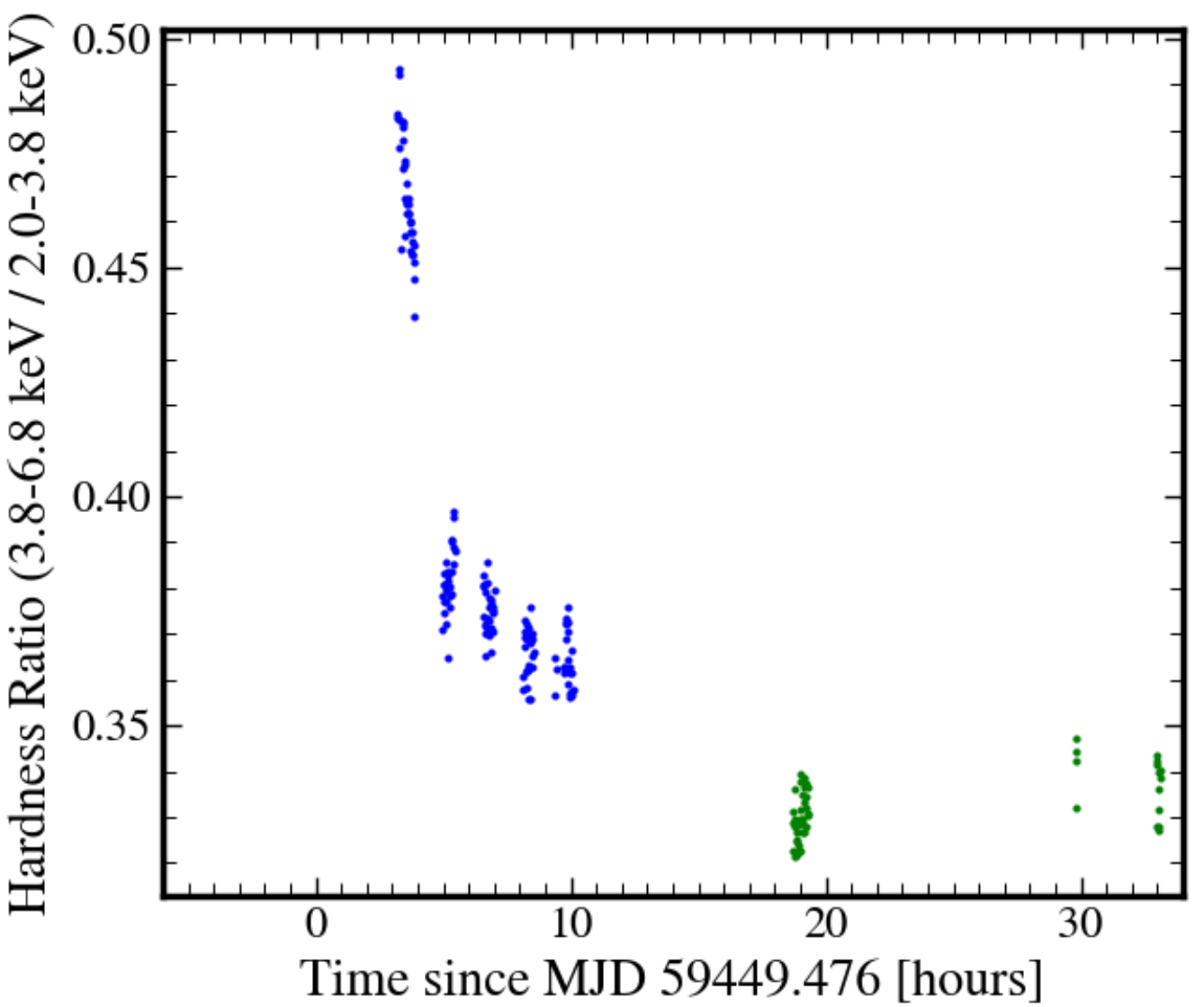}
    \caption{Hardness ratio of the superburst. We show the NICER hardness ratio between 3.8--6.8 and 2.0--3.8 keV. Each blue and green point represents a 64 s segment of data of NICER ObsID 05 and NICER ObsID 06, respectively.}
    \label{fig:2}
\end{figure}

The three NICER observations were processed following the standard procedure using HEASOFT version 6.32.1 and the NICER Data Analysis Software. The default filtering criteria were applied to obtain the cleaned event data. The light curves in the energy range 0.5--10 keV were extracted using \texttt{nicerl3-lc}; see Fig. \ref{fig:1}. The count rate of ObsIDs 02 and 06 remained constant at $\approx2935$ and $\approx1977~{\rm counts~s^{-1}}$, respectively. However, during the superburst decay phase, the light curve from ObsID 05 increased from 1080 to 2204 ${\rm counts~s^{-1}}$ over 6.89 hr. The hardness ratio between 2.0--3.8 and 3.8--6.8 keV showed a gradual decline from 0.49 to 0.36 in ObsID 05 and a relatively stable 0.33 in ObsID 06, as shown in Fig. \ref{fig:2}, suggesting spectral state variation.

\begin{figure}[h]
    \centering
    \includegraphics[width=1\linewidth]{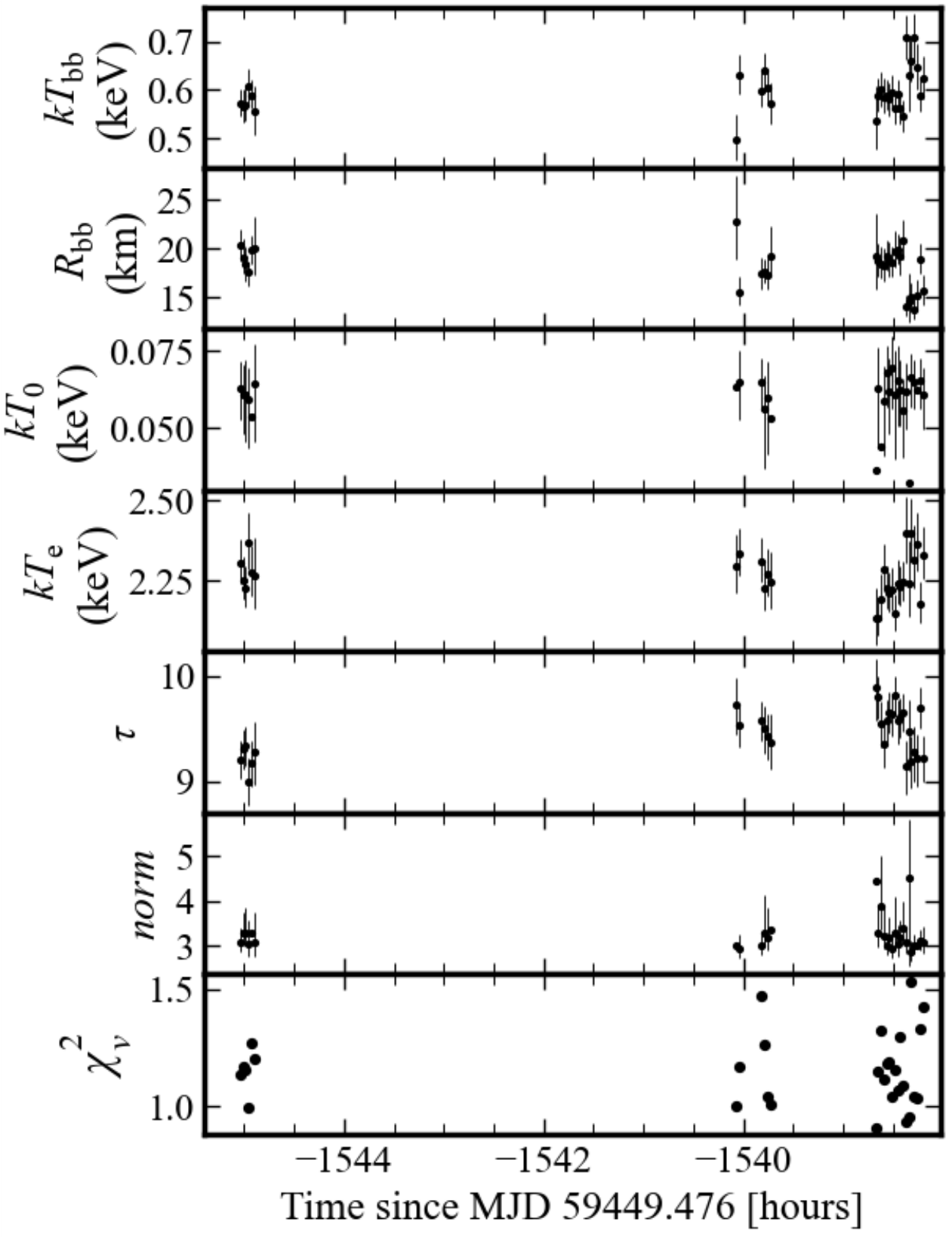}
    \caption{The spectral parameters of the persistent emission from ObsID 02. From top to bottom, the blackbody temperature and the blackbody radius, which were calculated using a distance of 8.4 kpc; the temperatures of the seed photons; the temperatures of the hot electrons; the optical thickness of the electron slab; the normalization; and the goodness of fit per dof, $\chi^{2}_{\nu}$, are shown. }
    \label{fig:33}
\end{figure}

\section{The spectral analysis} \label{sec:3}
We extracted the NICER spectra, ancillary response files, and response matrix files (RMFs) using \texttt{nicerl3-spect}, and the 3C50 background spectra were also produced simultaneously \citep{2022AJ....163..130R}. We studied the time-resolved persistent spectra in Sect.~\ref{sec:per} and the burst spectra in Sect.~\ref{sec:burst} using XSPEC v12.13.1 \citep{1996ASPC..101...17A}. We described the interstellar absorption using the Tübingen-Boulder model \citep[\texttt{TBabs};][]{Wilms_2000} with abundances from \citet{Wilms_2000}. 
The bolometric fluxes were estimated in the energy range of 0.5--100 keV by using the cflux tool. The uncertainties were reported at a $1\sigma$ confidence level.

\subsection{Persistent emissions}\label{sec:per}

We fitted the time-resolved persistent spectra from ObsIDs 02 and 06 observed by NICER. Each spectrum lasts 100~s.
Optimal binning for the persistent spectra was applied using \texttt{ftgrouppha} as recommended by the NICER team.

Previous spectroscopic studies of 4U 1820--30 have found that the persistent X-ray continuum can be well described with a model consisting of an absorbed blackbody plus a Comptonization component, \texttt{Tbabs*(bbodyrad+compTT)} \citep{2024A&A...683A..93Y}. 
We also adopted this model to fit the persistent spectra.
Model parameters include the blackbody temperature, $kT_{\rm bb}$, and its normalization, $K = (R_{\rm bb}/D_{\rm 10~kpc})^2$, where $D_{\rm 10~kpc}$ is the distance to the source in units of 10~kpc, for \texttt{bbodyrad};
the temperatures of the seed photons and hot electrons, $kT_{\rm 0}$ and $kT_{\rm e}$, respectively; the optical thickness of the electron slab, $\tau$, and the normalization, for \texttt{compTT};
and the equivalent hydrogen column, $N_{\rm H}$, for \texttt{Tbabs}. 
The \texttt{compTT} geometry was set as disk.
The absorption column density was initially set as a free parameter but showed negligible variation, so it was fixed at the mean values, $N_{\rm H}= 2.5 \times 10^{21}~ {\rm cm^{-2}}$ for ObsID 02 and $N_{\rm H}= 2.2 \times 10^{21}~{\rm cm^{-2}}$ for ObsID 06.
The obtained $N_{\rm H}$ are well consistent with the range of $1.9\textbf{--}2.5\times10^{21}~ {\rm cm^{-2}}$ reported by \citet{2024ApJ...975...67J} and \citet{2024A&A...683A..93Y}.

\begin{figure}[h]
    \centering
    \includegraphics[width=1\linewidth]{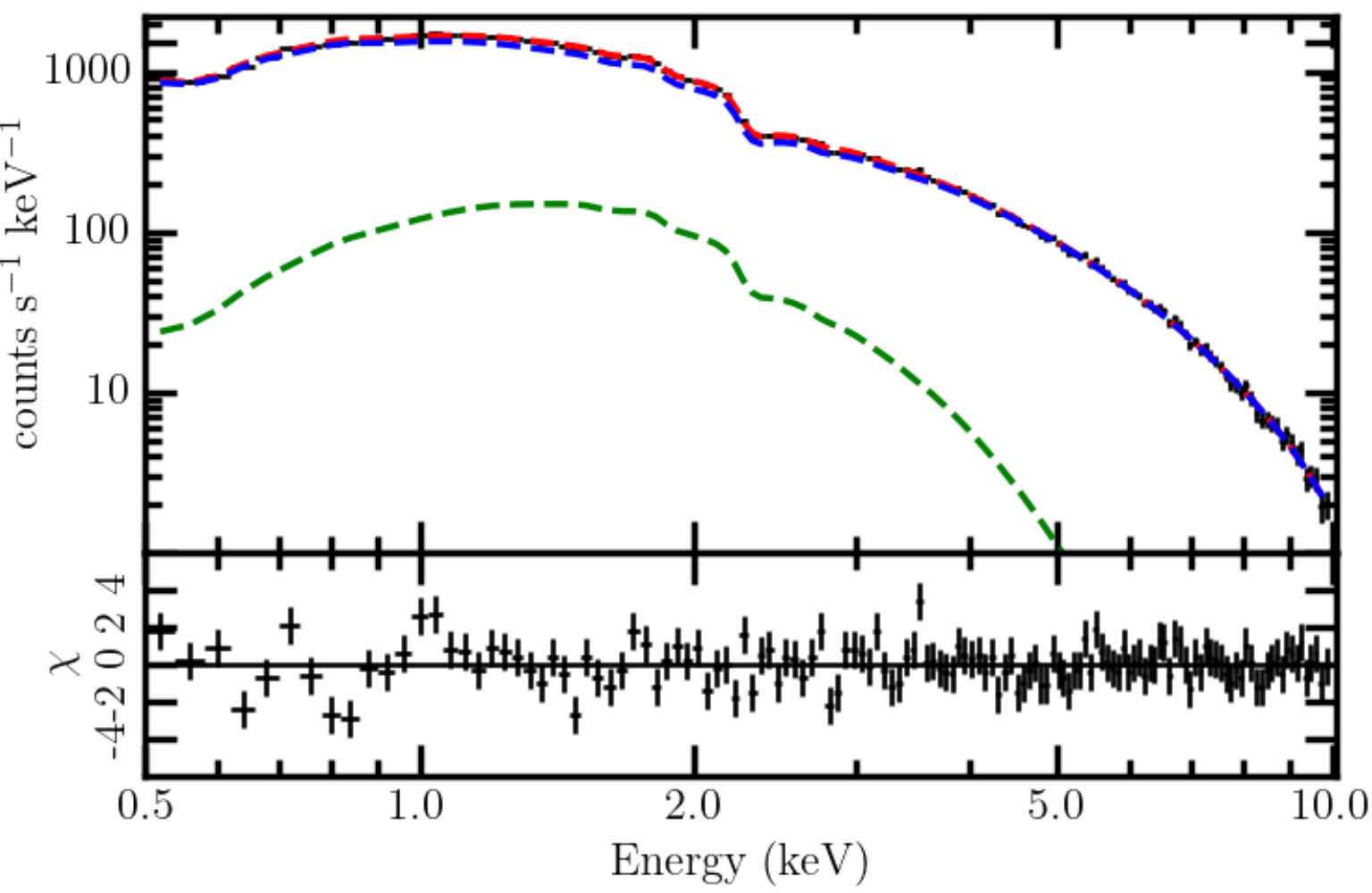}
    \caption{The absorbed best-fit persistent spectrum obtained between MJD 59385.10133--59385.10249 and the residuals from NICER ObsID 02 in 0.5--10 keV. The spectrum is fitted with the model \texttt{Tbabs*(bbodyrad+compTT)}. The red solid, blue dashed–dotted, and green dashed lines represent the best-fit model, the \texttt{compTT}, and the \texttt{blackbody} component, respectively.}
    \label{fig:44}
\end{figure}

This model can fit all persistent spectra well for the $\chi^2$ per degree of freedom (dof), $\chi^2_\nu<1.5$. The best-fit parameters from ObsID 02 are shown in Fig.~\ref{fig:33}. The persistent spectra showed a quite similar shape, with $kT_{\rm bb}\approx0.60$~keV, $R_{\rm bb}\approx18.12$~km, $kT_{\rm 0}\approx0.059$~keV, $kT_{\rm e}\approx2.26$~keV, $\tau\approx9.46$, and a normalization of $\approx3.24$, indicating stable persistent emissions. We show one of the best-fit spectrum and residuals in Fig.~\ref{fig:44} as an example.
 
\begin{figure}[h]
    \centering
    \includegraphics[width=1\linewidth]{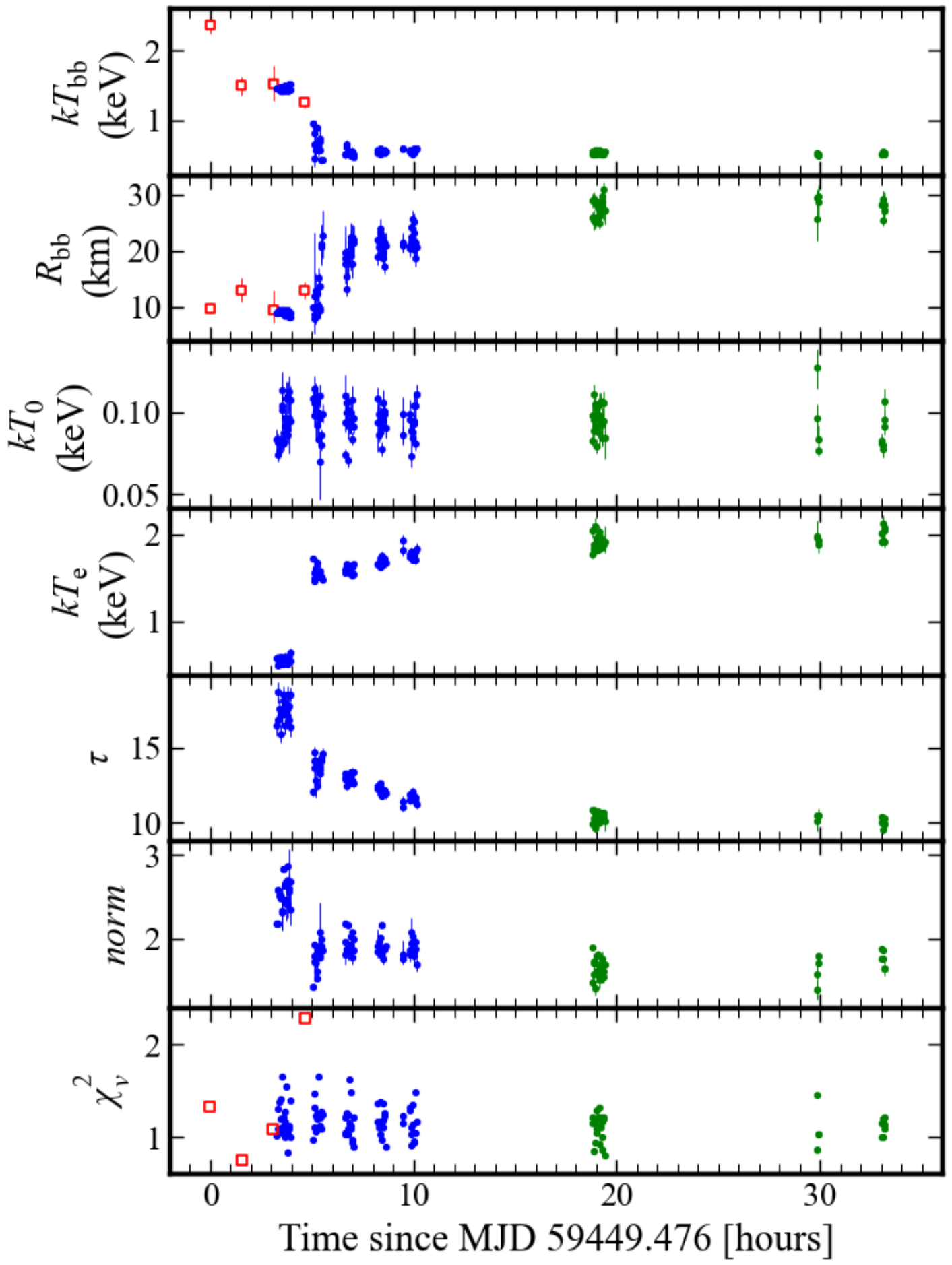}
    \caption{The spectral parameters of the superburst
in 4U 1820--30 evolve with time. From top to bottom, the blackbody temperature and blackbody radius, which were calculated using a distance of 8.4 kpc; the temperatures of the seed photons; the temperatures of the hot electrons; the optical thickness of the electron slab; the normalization;
and the goodness of fit per dof, $\chi^{2}_{\nu}$ are shown. The red squares are the data from MAXI, blue points are for ObsID 05 and green points are for ObsID 06.}
    \label{fig:55}
\end{figure}

\subsection{Time-resolved Burst Spectroscopy}\label{sec:burst}

We performed the time-resolved burst spectral analysis for MAXI and NICER observations separately. For the MAXI data, we downloaded four time-resolved burst spectra and the corresponding background spectra and RMFs with an exposure time of 120 s from MAXI/GSC.\footnote{\href{http://maxi.riken.jp/mxondem/}{http://maxi.riken.jp/mxondem/}} For the NICER observation, we extracted the time-resolved burst spectra with an exposure time of 100~s from the cleaned event file in ObsID 05. Only the instrumental background is subtracted. Therefore, the spectra contributed to the combination of the burst and persistent emissions. For the MAXI burst spectra, we first tried the model \texttt{TBabs(bbodyrad+compTT)}; however, the contribution of \texttt{compTT} can be neglected. We obtained the best-fitting parameters of the blackbody component and set the upper limit bolometric flux of the \texttt{compTT} component. The last MAXI spectrum did not fit well because of its low signal-to-noise ratio. 

For the NICER burst spectra, we used the model \texttt{TBabs(bbodyrad+compTT)gabs} to fit the burst spectra, where the model \texttt{gabs} is involved to account for the absorption line in some burst spectra; see Sect.~\ref{sec:line1} for more details. The absorption column density did not vary significantly; therefore, it was fixed at the mean value of $N_{\rm H}= 2.2\times 10^{21}~ {\rm cm^{-2}}$. 
The model can well describe the burst spectra for $\chi^{2}_{\nu}<1.5$, and no apparent features are shown in the residuals.

\begin{figure}[h]
    \centering
    \includegraphics[width=1\linewidth]{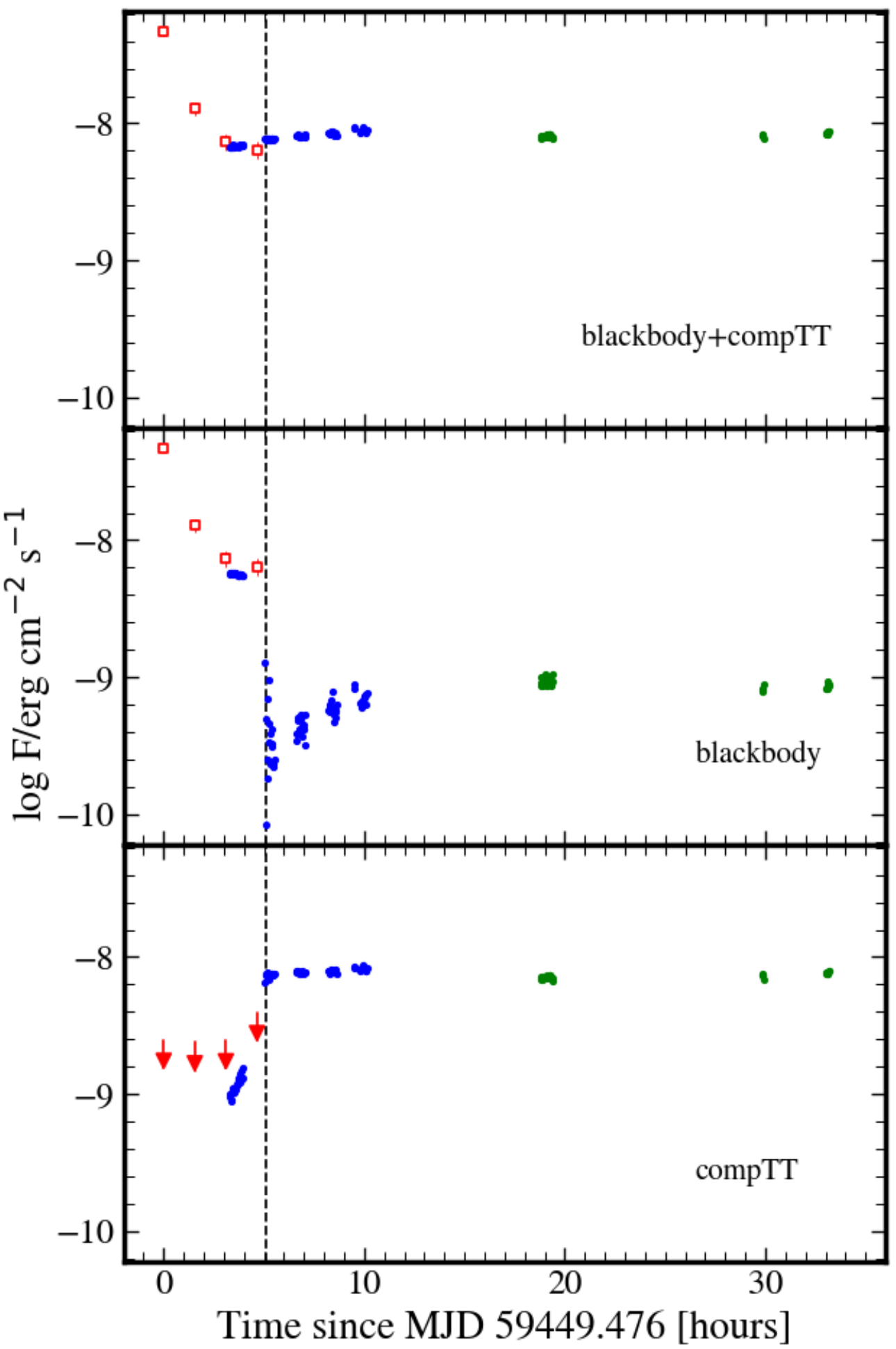}
    \caption{ 
    From top to bottom, the evolution of the flux represented as the total, the blackbody, and the Comptonization flux from ObsIDs 05 and 06. The dotted line marks the moment of the lowest blackbody flux.}
    \label{fig:66}
\end{figure}

We show the best-fitting parameters and $\chi^2_\nu$ of the MAXI and NICER spectra in Fig.~\ref{fig:55}. 
For the first 5~hr, the blackbody temperature slowly decreased from $\approx2.38$ to $\approx$ 1.26~keV implying the burst cooling. The blackbody radius stayed around 10~km, implying that the bursts were emitted over the whole NS surface. In next 0.11 hr, the blackbody temperature rapidly decreased from 0.96 to 0.45~keV, and the blackbody radius increased from 10 to 11.85~km and stabilized around 27.64 km in ObsID 06, same as the results from ObsID 02. The size of the blackbody component of all spectra is larger than the typical radius of the NS, which implies its emission from the boundary layer \citep{PS01,Gilfanov03}. 
The seed photon temperature $kT_{\rm 0}$ of the \texttt{compTT} component is mildly variable, and its value is around 0.094 keV. The temperatures of hot electrons $kT_{\rm e}$ gradually increased from 0.49 to 1.92 keV, while $\tau$ and normalization gradually decreased from 18.71 to 11.27 and from 2.83 to 1.69, respectively, close to the preburst levels. 

\begin{figure}[h]
    \centering
    \includegraphics[width=1\linewidth]{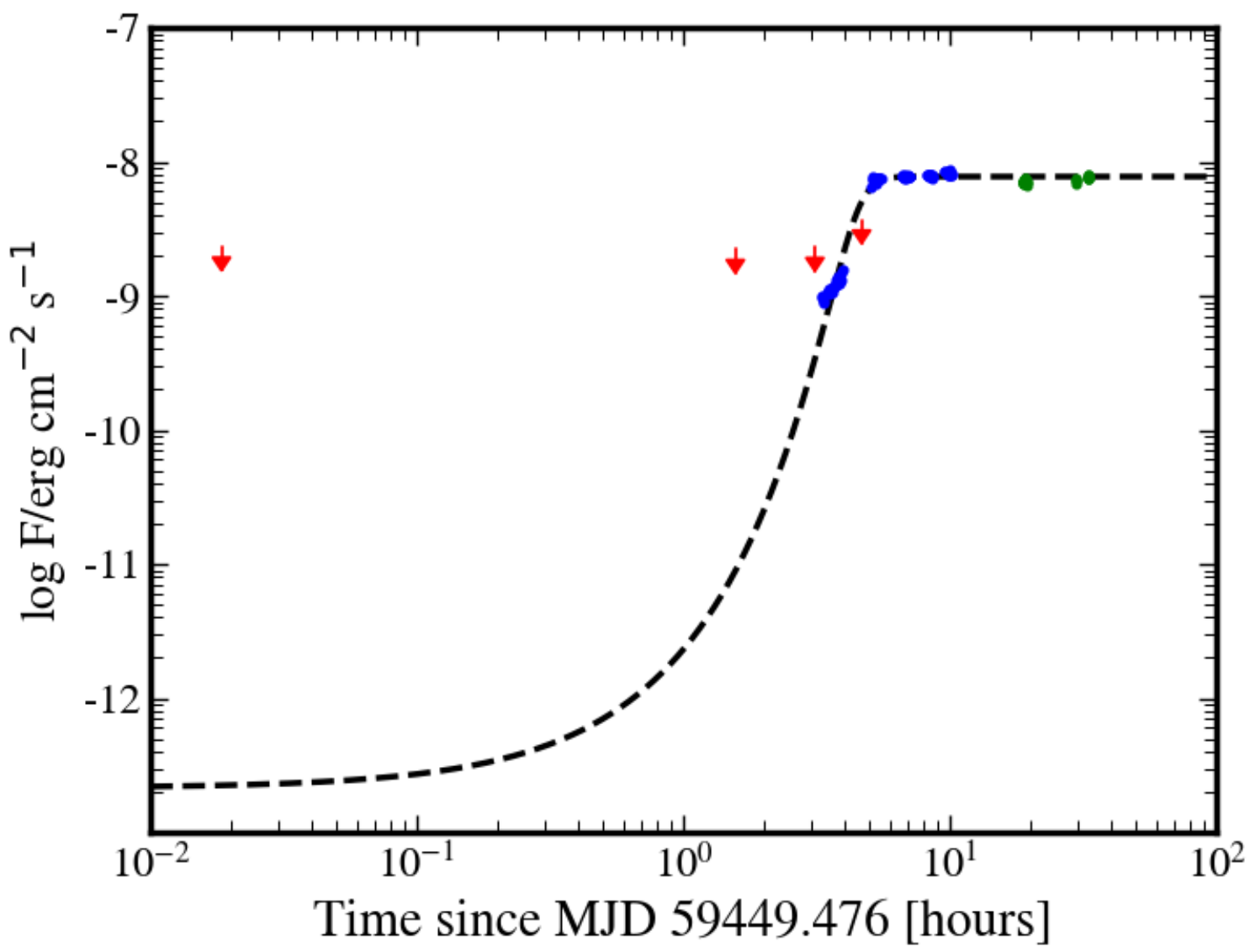}
    \caption{Evolution of the Comptonization flux fitted by the sigmoid model $f(t) = F/(1 + e^{-k(t - t_0)})$. The red, blue, and green points represent the data from MAXI and ObsIDs 05 and 06, respectively.}
    \label{fig:77}
\end{figure}

The total, blackbody, and Comptonization fluxes from MAXI and NICER are shown in Fig.~\ref{fig:66}.
The peak blackbody flux was $4.6\times 10^{-8}~{\rm erg~s^{-1}~cm^{-2}}$, less than the value observed with $RXTE/PCA$ from 4U 1820--30, $F_{\rm Edd}\approx 5.4\times10^{-8}~{\rm erg~s^{-1}~cm^{-2}}$ \citep{2008ApJS..179..360G}. However, the peak of the superburst might be missed, and the PRE phase in this superburst cannot be ruled out. 
For the first 5.11~hr since the burst trigger, the blackbody flux gradually dropped from $4.6\times 10^{-8}$ to $8.40\times 10^{-11}~{\rm erg~s^{-1}~cm^{-2}}$, and in the next 5.03 hr, the blackbody flux increased to the preburst level of $9.02\times 10^{-10}~{\rm erg~s^{-1}~cm^{-2}}$.
As the X-ray burst emission decayed, the Comptonization flux increased from $8.95\times 10^{-10}~{\rm erg~s^{-1}~cm^{-2}}$ to the preburst level, $7.29\times 10^{-9}~{\rm erg~s^{-1}~cm^{-2}}$. 
The evolution of the Comptonization flux can be well fitted by the sigmoid function $f(t) = F/(1 + e^{-k(t - t_0)})$, which describes an S-shaped evolution, capturing the initial rapid growth, the transitional behavior, and the final asymptotic approach to a saturation value. We obtained the best-fit parameters, the maximum value of the Comptonization flux, $F=(7.74\pm0.01)\times10^{-9}~{\rm erg~s^{-1}~cm^{-2}}$, the logistic growth rate, $k=2.38\pm0.01~{\rm hr^{-1}}$, the midpoint of the sigmoid and $t_0=4.41\pm0.01$~hr; see Fig.~\ref{fig:77}. If the rise time scale, $t_{\rm rise}$, is defined as the time interval over which the function rises from 10\% to 90\% of its maximum value, $F$, it can be expressed as $t_{\rm rise}=2\ln 9/k\approx4.4/k\approx1.8~{\rm hr}$. We estimated the Comptonization flux to be approximately $2.2\times 10^{-13}~{\rm erg~s^{-1}~cm^{-2}}$ around the peak of the superburst. This value is considered to be the lower limit of the persistent flux. Accordingly, we report the range of persistent flux as  $2.2\times 10^{-13}\text{--}9\times10^{-10}~{\rm erg~s^{-1}~cm^{-2}}$ reflecting an almost complete quenching of the persistent emission. 

\subsection{The absorption line}\label{sec:line1}
\begin{figure}[h]
    \centering
    \includegraphics[width=1\linewidth]{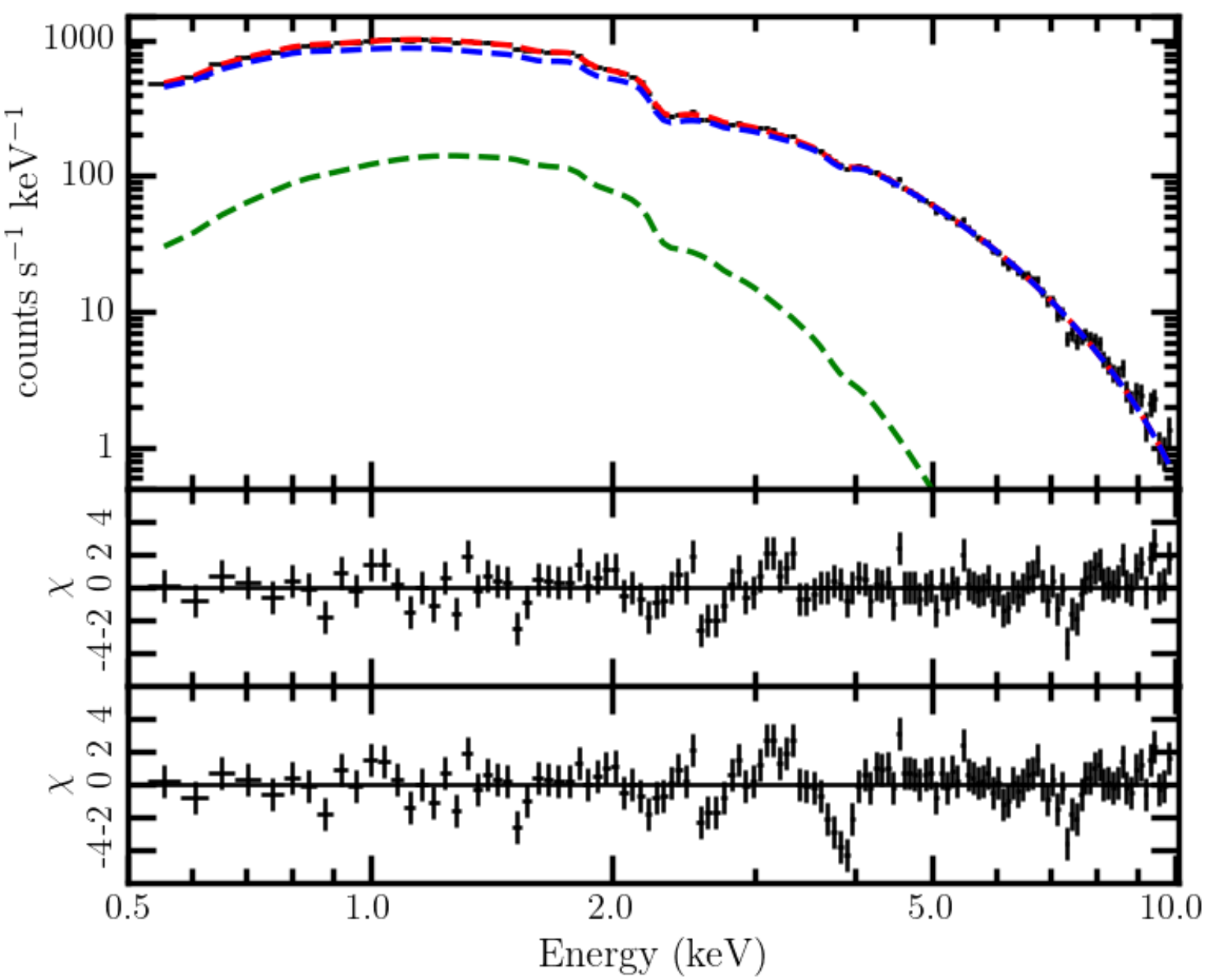}

    \caption{The burst spectrum obtained between MJD 59449.76493--59449.76609 and its best-fitting model from ObsID 05. In the top panel, we show the folded spectrum and the model \texttt{TBabs(bbodyrad+compTT)gabs}, while the blue dashed line, the green dashed line, and the red solid line represent \texttt{compTT}, \texttt{bbodyrad}, and the best-fit model, respectively. The residuals with and without absorption lines are plotted in the middle and bottom panels, respectively.}
    \label{fig:88}
\end{figure}

\begin{figure}[h]
    \centering
    \includegraphics[width=1\linewidth]{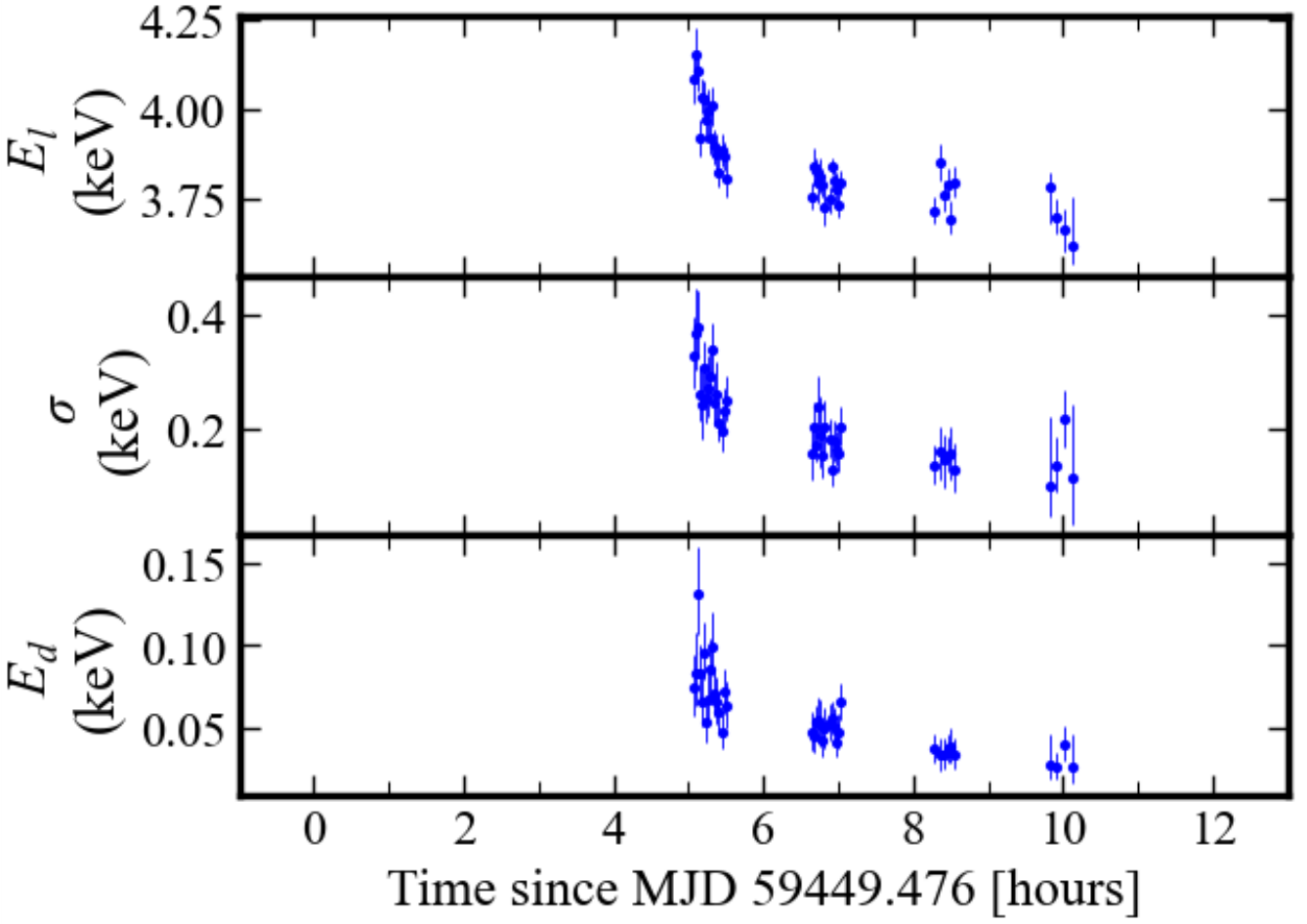}
    \caption{From top to bottom, we show the evolution of the absorption line energy, $E_l$, width, $\sigma$, and depth, $E_d$.
}
    \label{fig:99}
\end{figure}

From the time-resolved burst spectral analysis, we found an absorption line around 3.8 keV in some spectra of ObsID 05.
As an example, we tried the model \texttt{TBabs(bbodyrad+compTT)} to fit one burst spectrum in ObsID 05 and obtained $\chi^2_\nu=1.67$ for 125 dof, and there is a significant absorption feature in the residual around 3.84 keV. We added \texttt{gabs} to the model and found the fit improved to $\chi^2_\nu=1.12$ for 129 dof; see Fig.~\ref{fig:88}. 
The necessity of incorporating a Gaussian component was evaluated using the \texttt{simftest} command in Xspec. We run 1000 simulations and only keep the line with a significance higher than 3$\sigma$. We found that the absorption line was significant in 5--10 hr since the burst trigger; see Fig.~\ref{fig:99}. During this period, 39 out of 59 burst spectra were fitted with the addition of the gabs model. The line energy, $E_l$, is decreased from 4.15 to 3.62 keV accompanied by the reduction of the line width, $\sigma$, and depth, $E_d$, from 0.38 to 0.09 keV and from 0.13 to 0.03 keV, respectively.

\section{Discussion and Summary}\label{sec:4}

In this work, we reported the superburst from 4U 1820--30 observed by NICER and MAXI in 2021 August. We performed the spectral analysis of the superburst and persistent emissions. From the burst flux, we constrain the burst fuel in Sect.~\ref{sec:cooling}. We found the variation of persistent flux during the superburst, which is discussed in Sect.~\ref{sec:comp_flux}. We also detected the absorption line around 4.0~keV and explain its origin and variation in Sect.~\ref{sec:line}.

\subsection{Constraints on the Burst Parameters from the Superburst Cooling Flux}
\label{sec:cooling}

\begin{figure}[h]
    \centering
    \includegraphics[width=1\linewidth]{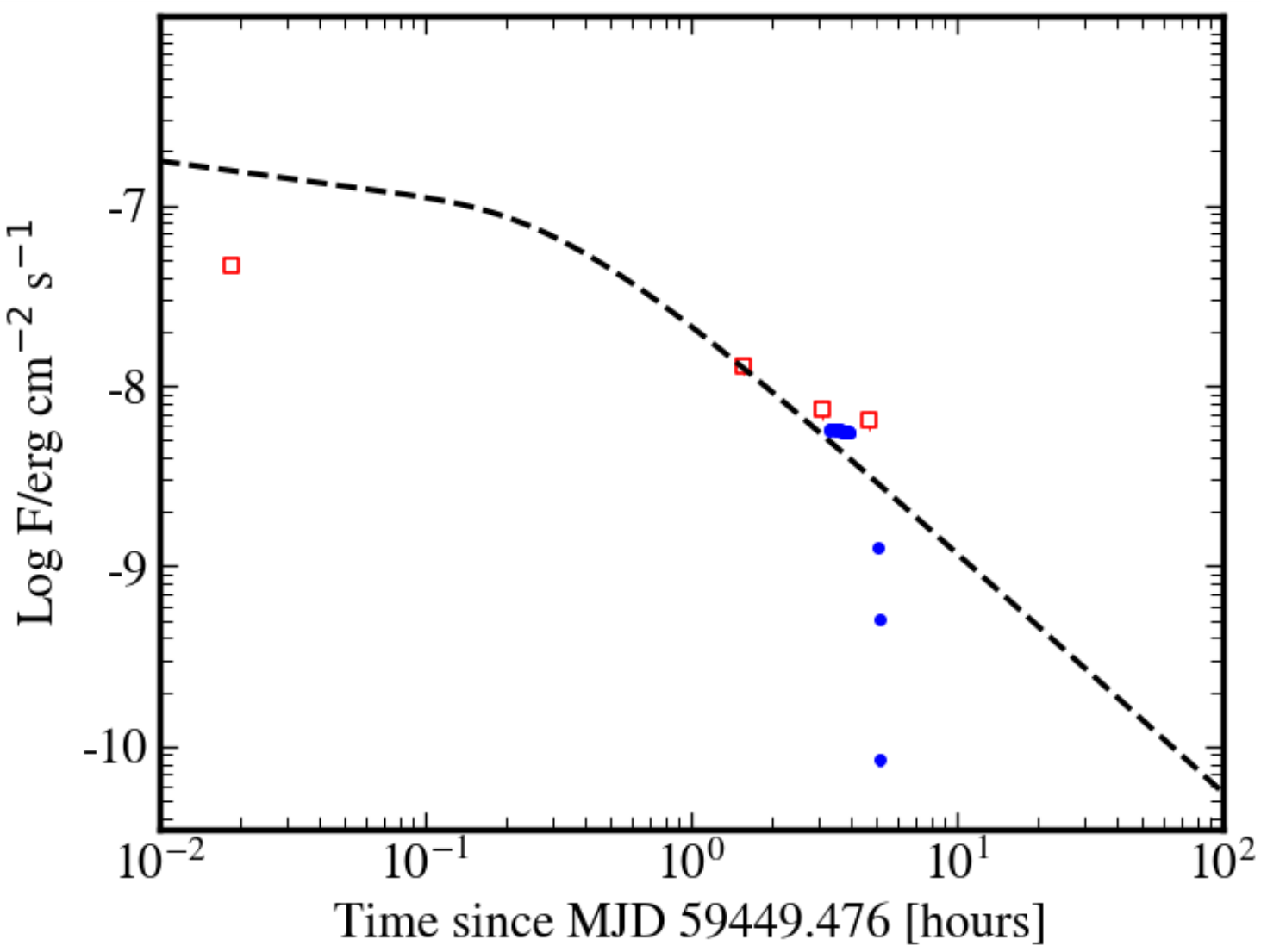}
    \caption{The decay of the superburst blackbody flux fitted by the model from
\citet{Cumming_2004}. The red and blue points represent the data from
MAXI and NICER, respectively.}
    \label{fig:10}
\end{figure}

For the first 5.11 hr after the superburst trigger, the blackbody temperature dropped with an invariant blackbody radius, indicating the superburst cooling on the NS surface. We fitted the decay of the superburst blackbody flux with the analytic expression provided by \citet{Cumming_2004} and \citet{2006ApJ...646..429C}, which depends on the energy release per unit mass, $E_{17}$, in units of 10$^{17}~{\rm erg~g^{-1}}$; and the ignition column depth $y_{12}$ in units of 10$^{12}~{\rm g ~cm^{-2}}$; see Fig. \ref{fig:10}. Even though the $\chi_{\nu}^2$ is large, the evolution of the superburst flux is broadly consistent with the model prediction. 
We obtained $E_{\rm 17}\sim2.37$ and $y_{\rm 12}\sim0.29$. We then calculated the burst fluence from the estimated ignition column depth $y_{\rm ign}$,
\begin{align}
    f_{\rm b}=\frac{4\pi y_{\rm ign}R^{2}_{\rm NS}Q_{\rm nuc}}{4\pi d^{2}(1+z)},
\end{align}
where $R_{\rm NS}= 10$~km, $Q_{\rm nuc} \approx 1.31 ~{\rm MeV~nucleon^{-1}}$, the source distance $d = 8.4$~kpc and the gravitational redshift on the NS surface $z=0.31$ for the NS mass of $M_{\rm NS}=1.4 M_{\odot}$.
We obtained the burst fluence $f_{\rm b}\approx 4.1\times 10^{-4}~{\rm erg~cm^{-2}}$ and the total burst energy release of $\approx3.5\times10^{42}~{\rm erg}$. We then calculated the burst decay time $\tau = f_{\rm b}/F_{\rm peak} \approx 2.5$~hr, which is higher than the exponential decay time of 0.97 hr obtained from the light curve. 
We determined the local accretion rate, 
\begin{align}
\label{Eq:6}
    \dot{m} & =\frac{L_{\rm per}(1 + z)}{4\pi R^{2}_{\rm NS}(GM_{\rm NS}/R_{\rm NS})}\notag \\
    & \approx 4.7 \times 10^{3} \left(\frac{F_{\rm per}}{10^{-9} \; {\rm erg} \; {\rm cm^{-2}} \; {\rm s^{-1}}}\right)\left(\frac{D}{8.4 \; {\rm kpc}}\right)^{2} \\
    & \times \left(\frac{M_{\rm NS}}{1.4M_{\odot}}\right)^{-1}\left(\frac{1+z}{1.31}\right)\left(\frac{R_{\rm NS}}{\rm 10\;km}\right)^{-1} {\rm g} \;{\rm cm^{-2}} \;{\rm s^{-1}}\notag,
\end{align}
 where $F_{\rm per}=8.2\times10^{-9}~{\rm erg~s^{-1}~cm^{-2}}$ is the persistent flux of ObsID 06 after the superburst. We obtained $\dot{m}=3.88\times 10^{4}~{\rm g~cm^{-2}~s^{-1}}$. The predicted recurrence time is calculated via the relation $\Delta t_{\rm rec} = y_{\rm ign}(1 + z)/\dot{m}\approx0.31$~yr. 
The observed recurrence time, $\Delta t_{\rm rec}\approx0.25$~yr, of two superbursts observed in 2021, 
which is close to the predicted value accounting for the variation of persistent flux during two superbursts.
Based on the spectral analysis, which yields consistent results, we propose that the superburst duration of 2.5~hr is more accurate than the 0.97~hr derived solely from the MAXI light curve.

\subsection{Implications of the blackbody and Componization flux evolution}\label{sec:comp_flux}

The interactions between the normal X-ray burst and the accretion disk have been studied in many aspects \citep[see, e.g.,][]{2018SSRv..214...15D,2018ApJ...867L..28F}. The burst radiation can remove the material of the inner accretion disk to form outflow \citep{2018SSRv..214...15D,Russell24} or drag the matter onto the NS surface due to the Poynting–Robertson effect \citep{Walker92,2020NatAs...4..541F,Zhao22,2023MNRAS.526.1388S}. Based on numerical simulations, the inner accretion disk shows signs of recovery after the burst, and the timescale of the full recovery may be comparable to or longer than the duration of the burst \citep{2020NatAs...4..541F, Speicher24}. From observations, the Poynting–Robertson drag effect and disk reflection features have been reported in superbursts \citep{2014ApJ...789..121K}. However, the simulation of the interactions between superburst and accretion disk is still lacking. Considering the total energy release of $3.5\times10^{42}~{\rm erg}$ of the studied superburst from 4U 1820--30, it is expected that the disk can be strongly affected.

During the cooling of the superburst, the Componization flux increased from $8.95\times 10^{-10}~{\rm erg~s^{-1}~cm^{-2}}$, i.e., $1.8\%L_{\rm Edd}$, to the pre- and postsuperburst levels, i.e., enhanced by a factor of 10, within 6.89 hr. Based on the NICER observations for 4U 1820--30, the bolometric persistent flux was in the range of $(2.5-15)\times 10^{-10}~{\rm erg~s^{-1}~cm^{-2}}$, higher than the lowest measured flux during this superburst \citep{2024A&A...683A..93Y}. Due to the lack of NICER data at the early stage of the superburst, we cannot directly measure the Componization flux during the superburst peak. However, the evolution of the Componization flux can be modeled by a sigmoid function, predicting a lower limit of $2.2\times 10^{-13}~{\rm erg~s^{-1}~cm^{-2}}$, while the actual flux could be orders of magnitude higher. Considering that the increase of persistent flux occurred during the cooling of a superburst, we propose that the variation of Componization flux was directly caused by the superburst itself. 

The burst radiation can significantly affect the inner accretion disk by expelling material through radiation pressure or emptying the inner accretion disk through an enhanced accretion flow via the Poynting–Robertson effect. To study the burst radiation pressure effect, we compare the superburst total energy release to the gravitational potential energy of the inner accretion material. Following the model of \citet{2005ApJ...626..364B}, the disk gravitational potential energy can be calculated from the disk surface density,
\begin{align}
\label{eq:3}
    \Sigma = 2.7\times10^{5}\alpha^{-4/5}\left(\frac{M_{\rm NS}}{1.4M_\odot}\right)^{-2/5}
\left(\frac{\dot{M}}{7.7\times10^{17}~{\rm g~s^{-1}}}\right)^{3/5}\notag\\
\times\left(\frac{D}{8.4 ~{\rm kpc}}\right)^{6/5}\left(\frac{R}{r_g}\right)^{-3/5}J(R)^{3/5}~{\rm g~cm^{-2}},
\end{align}
where $R$ denotes the outer radius of the accretion disk, $J(R)=1-(6r_g/R)^{1/2}$, $r_g=GM_{\rm NS}/c^2$, and $\alpha=0.1$ is the Shakura–Sunyaev viscosity parameter. The accretion rate is $\dot{M}=4\pi D^2F/\eta c^2=7.7\times10^{17}(D/8.4~{\rm kpc})^2~{\rm g~s^{-1}}$, where the persistent bolometric flux is $8.2\times10^{-9}~{\rm erg~s^{-1}~cm^{-2}}$, and the accretion efficiency $\eta$ is assumed to be 0.1. Based on this model, we estimated the gravitational potential energy, $\sim6\times10^{39}~{\rm erg}$, of the disk material located between $6r_g$ and $1000r_g$ under the Schwarzschild metric for an NS with a mass of $1.4M_\odot$ and radius of 10~km. If the peak flux of the superburst reached the Eddington limit, the energy release of the superburst, $3.5\times10^{42}~{\rm erg}$, exceeds the calculated potential energy, which can remove a significant portion of the accretion disk material. If the peak flux of the superburst was sub-Eddington, the Poynting–Robertson effect could drag the inner disk material onto the NS surface \citep{2005ApJ...626..364B, 2020NatAs...4..541F}. Subsequently, the burst radiation could prevent the inward migration of the accretion disk material \citep{2005ApJ...626..364B,Zand11}. As a consequence, the persistent emission was almost quenched. 

As the superburst radiation decayed, the material refilled the inner accretion disk, allowing the persistent emission to recover to its preburst level. The observed rise time of the Comptonization flux, $t_{\rm rise}\approx1.8~{\rm hr}$ (see Sect.~\ref{sec:burst}), can be interpreted as the viscous timescale for the inner accretion disk to refill, as described by \citep{2005ApJ...626..364B}
\begin{align}
    t_{\rm visc} = 0.35\alpha^{-4/5}\left(\frac{M_{\rm NS}}{1.4M_\odot}\right)^{8/5}
\left(\frac{\dot{M}}{7.7\times10^{17}~{\rm g~s^{-1}}}\right)^{-2/5}\notag\\
\times\left(\frac{D}{8.4 ~{\rm kpc}}\right)^{-4/5}\left(\frac{R}{1000r_g}\right)^{7/5}J(R)^{-2/5}~{\rm hr}.
\end{align}
If the same parameters are adopted as for Equation~(\ref{eq:3}), the recovery of the accretion disk from $1000r_g$ will take around $2.3~{\rm hr}$. Note that $\alpha$ can be varied between $\sim0.1$ and 0.4 \citep{King07}, and our estimation is close between the accretion recovery timescale and the rise time of the Comptonization flux. Particularly, if $\alpha\approx0.135$ is used, we can obtain the viscous timescale of around 1.8 hr, more closely matching the observed value. We conclude that the superburst radiation pressure or the Poynting–Robertson drag initially emptied the inner accretion disk, leading to quenching of persistent emission. As the superburst faded, accretion resumed on a reasonable viscous timescale, ultimately restoring the disk to its preburst state.

\subsection{Absorption Line}\label{sec:line}

Atomic features have been observed in many X-ray bursts, such as emission and absorption lines \citep{Degenaar13,Bult21,Lu24} and absorption edges \citep{Ballantyne04,Zand10,Kajava17,Li18}. If one of them can be confidently confirmed from the NS surface, the observed energy of the feature should be redshifted due to the strong gravitational field of the NS. Therefore, by comparing the observed energy to its rest-frame value, it can be used to measure the NS mass and radius ratio and constrain the NS equation of state \citep{Li18}.

Several spectral features from 4U 1820--30 were observed by NICER in the PRE phases of X-ray bursts, including an emission line around 1 keV (Fe or Ne line) and absorption lines around 1.7 (Mg, Fe, or Cr line) and 3 keV (He-like line of $\rm S_{~\underline{XV}}$). All of them were likely produced in the PRE wind; however, the 1 keV emission line may also originate from the reflection of the burst emission off the inner accretion disk \citep{2019ApJ...878L..27S}. 

We discuss the origin and evolution of the absorption line in 4U 1820--30, which decreased from 4.15 to 3.62 keV. The absorption line emerged during the recovery phase of the persistent emission, when the burst emission was negligible, making it unlikely to be associated with the NS surface. We propose that during the superburst, the ashes of unstable carbon burning, which were mainly composed of Si, S, and Ar, imprinted onto the accretion disk \citep{WB07}. To investigate the elements responsible for producing the observed line range, we consulted the NIST Atomic Spectra Database\footnote{https://www.nist.gov/pml/atomic-spectra-database} and identified Ar as the most likely origin. Therefore, we attribute the line to the $\rm Ar_{~\underline{XVIII}}$ originating from the inner accretion disk, which has a rest-frame energy of 4.15 keV. When the accretion disk was far from the NS surface, the absorption line was not shifted. As the accretion disk moved closer to the NS surface, the line was gravitationally redshifted by the NS. The lowest energy of 3.62 keV corresponds to a redshift factor of 0.146. For a typical NS mass of $1.4M_\odot$, the radius of the inner accretion disk is around 17~km, rather close to the NS surface. We note that this line did not appear during the superburst and fully recovered persistent phases. During the superburst, the accretion disk was located far from the NS; therefore, the solid angle of the accretion disk related to the NS is small, resulting in the nondetection of the absorption line. When the persistent emission fully recovered, the spectral fitting results indicated a high temperature in the inner accretion disk, leading to high ionization of the inner disk material. So, the absorption line disappeared and did not show again.

\section{Summary}\label{sec:5}
In this work, we detected a superburst from 4U 1820--30 using joint NICER and MAXI observations. These observations provide a unique opportunity to investigate the impact of a powerful thermonuclear burst on the accretion environment in UCXBs. Spectral analysis shows that the persistent emission was nearly quenched during the superburst, suggesting that the intense superburst radiation emptied the inner accretion disk. As the burst radiation diminished, the inner accretion was gradually restored to its preburst state. From the absorption line, we determined that the inner disk moved to a distance of around $17$~km from the NS surface. These observations highlight the profound impact that energetic superburst events can have on the structure and dynamics of the accretion disk in UCXB systems like 4U 1820--30, underscoring the need for further theoretical and observational investigations into this complex interaction.

\section*{Acknowledgments}
We appreciate the referee for the valuable comments and suggestions, which improved the manuscript. This work was supported by the Major Science and Technology Program of Xinjiang Uygur Autonomous Region (No. 2022A03013-3) and the National Natural Science Foundation of China (12103042, 12273030, U1938107). This work made use of data from the High Energy Astrophysics Science Archive Research Center (HEASARC), provided by NASA's Goddard Space Flight Center.

\bibliography{sample631}{}
\bibliographystyle{aasjournal}

\end{document}